\documentclass{wlscirep}
\usepackage{epsfig,graphicx,amsmath}
\usepackage{subfigure}
\title{ Single-photon multi-ports router based on the coupled cavity optomechanical system }

\author[1]{Xun Li}
\author[1]{Wen-Zhao Zhang}
\author[1,*]{Ling Zhou}
\affil[1]{School of Physics and Optoelectronic Technology, Dalian University of
Technology, Dalian, 116024, People's Republic of China}

\affil[*]{corresponding:zhlhxn@dlut.edu.cn}

\begin{abstract}
A scheme of single-photon multi-port router is put forward by coupling two
optomechanical cavities with waveguides. It is shown that the coupled two
optomechanical cavities can exhibit photon blockade effect, which
is generated from interference of three mode interaction. A single-photon travel along the system is
calculated. The results show that the single photon can be controlled  in the multi-port system
because of the radiation pressure, which should be  useful for constructing quantum network.
\end{abstract}

\begin{document}
\flushbottom
\maketitle
\flushbottom 

\thispagestyle{empty}

\section*{Introduction}

Quantum router to combine quantum channels with quantum nodes can create a
quantum network so as to distribute quantum information. Recently, many
theoretical proposals and experimental demonstrations of a quantum router
have been carried out in various systems, i.e., cavity QED system \cite
{Aoki2009,Hoi2011}, nanocavity arrays coupling with three-level system \cite
{Zhou2013,Lu2014a}, cavity electromechanical system \cite{Jiang2012b},
all-linear-optical scheme \cite{Lemr2013,Yuan2015,0295-5075-111-6-64005} and optomechanical systems \cite{Agarwal2012}.

Photon-blockade phenomenon resulted from nonlinearity allows only one photon existence, and the
second photon will be prohibited, which can be used to generate single photon source or to ensure a
single photon processing.  Cavity optomechanical systems, besides its potential application in
detecting gravity waves \cite{Ma2014,LIGOScientificCollaboration2009}, in studying
quantum-to-classical transitions \cite{Ghobadi2014a}, in performing high precision measurements
\cite {Zhang2015a,Barzanjeh2015,Kippenberg2013,Chen2013},in entanglement generation
\cite{Vitali2007a,Bai2016a} and preservation\cite{Cheng2016} and in processing quantum information
\cite{Zhang2015b,Li2015b,Kippenberg2013,Dalafi2013,Xu2015}, are of nonlinearity
\cite{Gong2009,Ludwig2012,Liao2015,Liu2015a,Flayac2015,Rabl2011,Wang2015}. But this nonlinear strength
proportional to $g^{2}/\omega _{m}$ is limited by the condition $g$ (the coupling strength of
radiation pressure) less than $ \omega _{m}$ (the frequency of the mechanical oscillator),
therefore, a lot of effort is devoted to enhance the nonlinearity, for instance, adding
atoms\cite{Zhou2011}, introducing quantum dot\cite{Tang2015}, using coupled cavity optomechanical
system \cite{Ludwig2012}. In Ref.\cite{Komar2013a}, the authors had put forward a approach where the
photons in the two optical modes can be resonantly exchanged by absorbing or emitting a phonon via
three-mode mixing so as to generate effective photon blockade not depending on the ratio
$g^{2}/\omega _{m}$.

In this paper, we put forward a scheme by coupling two cavity optomechanical
system. We show that our system can be effective equal to three-mode
interaction\cite{Komar2013a} and can exhibit photon blockade. Then we
construct four output ports by coupling wave guide to the
two-cavity-optomechanical system. Our research show that our system can work
as multiple output ports router under the assistant of mechanical mode,
which provide a potential application for the cavity optomechanical system
in multiple router.

\section*{Results}

In this part, we introduce the model, illustrate the photon-blockade effect
of this two-cavity-optomechanical waveguide coupled system and study the
transport of photons of waveguide under photon-blockade condition.

\subsection*{Model and effective interaction}

\begin{figure}[tbp]
\centering\includegraphics[width=15cm]{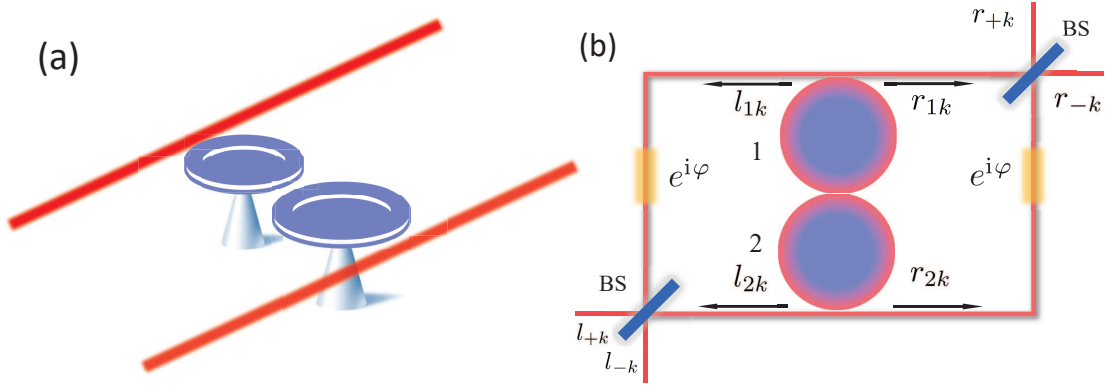}
\caption{Schematic configuration of the single-photon router. (a)The two
toroidal cavities with mechanical modes coupling to waveguide. (b)The four
ports router with quasi-mode. }
\label{figure_configure}
\end{figure}

We consider the two optomechanical cavities coupled with hopping coefficient
$J$ where the two cavities are driven by two laser beam with frequencies $
\omega _{L_{1}}$ and $\omega _{L_{2}}$ separately, and the two
optomechanical cavities are side-coupled to the fibres. The configuration of
the system is shown in Fig.\ref{figure_configure}a, which is similar with Ref. \cite{Chang2014}
where they utilized the two coupled whispering-gallery-mode (WGM)
microtoroids coupled to two tapered fibres to experimentally realize
parity--time-symmetric optics, but the mechanical modes are ignored. Taking
the mechanical modes into consideration, we write the Hamiltonian as

\begin{equation}
    \hat{H}=\hat{H}_{cav}+\hat{H}_{om}+\hat{H}_{f}  \label{H_totoal}
\end{equation}
with
\begin{eqnarray}
\hat{H}_{cav} &=&\omega _{1}\hat{a}_{1}^{\dagger }\hat{a}_{1}+\omega _{2}
\hat{a}_{2}^{\dagger }\hat{a}_{2}+J(\hat{a}_{1}^{\dagger }\hat{a}_{2}+\hat{a}
_{1}\hat{a}_{2}^{\dagger })+\sum_{j=1,2}\varepsilon _{j}(\hat{a}
_{j}e^{i\omega _{L_{j}}t}+\hat{a}_{j}^{\dagger }e^{-i\omega _{L_{j}}t}),
\notag \\
\hat{H}_{om} &=&\;\omega _{m1}\hat{b}_{1}^{\dagger }\hat{b}_{1}+\omega _{m2}
\hat{b}_{2}^{\dagger }\hat{b}_{2}+g\hat{a}_{1}^{\dagger }\hat{a}_{1}(\hat{b}
_{1}+\hat{b}_{1}^{\dagger })+g\hat{a}_{2}^{\dagger }\hat{a}_{2}(\hat{b}_{2}+
\hat{b}_{2}^{\dagger }),  \notag
\end{eqnarray}
where $\hat{H}_{cav}$ describes the free energy of the cavity, the exchange
energy between the two cavities with hopping strength $J$ as well as the two
classical driving laser with frequencies $\omega _{L_{1}}$ and $\omega
_{L_{2}}$ respectively, where $\hat{a}_{1}(\hat{a}_{2})$ and $\hat{a}
_{1}^{\dagger }(\hat{a}_{2}^{\dagger })$ represent the annihilation and
creation operators of cavity modes. $\hat{H}_{om}$ represents the energy of
the two mechanical oscillators and their coupling with the cavity fields
induced by radiation pressure, where the $\hat{b}_{1}(\hat{b}_{2})$ and $
\hat{b}_{1}^{\dagger }(\hat{b}_{2}^{\dagger })$ are annihilation and
creation operators of mechanical oscillators. The Hamiltonian $\hat{H}_{f}$
in {Eq.(\ref{H_totoal})} can be written as
\begin{equation}
\hat{H}_{f}=\sum_{O=r,l}\sum_{j=1,2}\int_{-\infty }^{+\infty }dk\left[
\omega _{k}\hat{O}_{jk}^{\dagger }\hat{O}_{jk}+i\xi (\hat{a}_{j}^{\dagger }
\hat{O}_{jk}-\hat{a}_{j}\hat{O}_{jk}^{\dagger })\right] ,
\end{equation}
which expresses the two cavity fields coupling with the fibres, where $\hat{O
}_{jk}$ ($j=1,2;\hat{O}=r,l)$ represents annihilation operators of the
fibres. In the frame rotating with $\hat{H}_{0}=\omega _{L_{1}}[\hat{a}
_{1}^{\dagger }\hat{a}_{1}+\sum_{\hat{O}=\hat{r},\hat{l}}\int_{-\infty
}^{+\infty }dk\hat{O}_{1k}^{\dagger }\hat{O}_{1k}]+\omega _{L_{2}}[\hat{a}
_{2}^{\dagger }\hat{a}_{2}+\sum_{\hat{O}=\hat{r},\hat{l}}\int_{-\infty
}^{+\infty }dk\hat{O}_{2k}^{\dagger }\hat{O}_{2k}]$ , we have
\begin{eqnarray}
\hat{H}_{cav}^{\prime } &=&\Delta (\hat{a}_{1}^{\dagger }\hat{a}_{1}+\hat{a}
_{2}^{\dagger }\hat{a}_{2})-J(\hat{a}_{1}^{\dagger }\hat{a}_{2}+\hat{a}_{1}
\hat{a}_{2}^{\dagger })+\sum_{j=1,2}\varepsilon _{j}(\hat{a}_{j}+\hat{a}
_{j}^{\dagger }),  \notag \\
\hat{H}_{om}^{\prime } &=&\omega _{m_{1}}\hat{b}_{1}^{\dagger }\hat{b}
_{1}+\omega _{m_{2}}\hat{b}_{2}^{\dagger }\hat{b}_{2}+g_{1}\hat{a}
_{1}^{\dagger }\hat{a}_{1}(\hat{b}_{1}+\hat{b}_{1}^{\dagger })+g_{2}\hat{a}
_{2}^{\dagger }\hat{a}_{2}(\hat{b}_{2}+\hat{b}_{2}^{\dagger }),
\end{eqnarray}
where for simplicity we assume $\omega _{1}=\omega _{2}$ and $\omega
_{L_{1}}=\omega _{L_{2}}$ so that $\Delta =\omega _{1}-\omega
_{L_{1}}=\omega _{2}-\omega _{L_{2}}$ , and
\begin{equation}
\hat{H}_{f}=\sum_{O=r,l}\sum_{j=1,2}\int_{-\infty }^{+\infty }dk\left[
\Delta _{k}O_{jk}^{\dagger }O_{jk}+i\xi (\hat{a}_{j}^{\dagger }O_{jk}-\hat{a}
_{j}O_{jk}^{\dagger })\right] ,
\end{equation}
where $\Delta _{k}=\omega _{k}-\omega _{L_{1}}=\omega
_{k}-\omega _{L_{2}}$. Now, we introduce the operators
\begin{equation*}
\hat{a}_{\pm }=\frac{1}{\sqrt{2}}(\hat{a}_{1}\pm \hat{a}_{2}),\hat{b}_{\pm }=
\frac{1}{\sqrt{2}}(\hat{b}_{1}\pm \hat{b}_{2}),
\end{equation*}
The Hamiltonian $\hat{H}_{s}=\hat{H}_{cav}+\hat{H}_{om}$ is of the form
\begin{eqnarray}
\hat{H}_{s} &=&\Delta _{+}\hat{a}_{+}^{\dagger }\hat{a}_{+}+\Delta _{-}\hat{a
}_{-}^{\dagger }\hat{a}_{-}+\varepsilon _{-}(\hat{a}_{-}^{\dagger }+\hat{a}
_{-})+\varepsilon _{+}(\hat{a}_{+}^{\dagger }+\hat{a}_{+})+\omega _{m}\hat{b}
_{+}^{\dagger }\hat{b}_{+}+\omega _{m}\hat{b}_{-}^{\dagger }\hat{b}_{-}
\notag \\
&&+\frac{g}{\sqrt{2}}(\hat{b}_{+}+\hat{b}_{+}^{\dagger })(\hat{a}
_{+}^{\dagger }\hat{a}_{+}+\hat{a}_{-}^{\dagger }\hat{a}_{-})+\frac{g}{\sqrt{
2}}(\hat{b}_{-}+\hat{b}_{-}^{\dagger })(\hat{a}_{-}^{\dagger }\hat{a}_{+}+
\hat{a}_{+}^{\dagger }\hat{a}_{-}),  \label{eq-H_int3}
\end{eqnarray}
where we have assume $g_{1}=g_{2}=g$, $\Delta _{\pm }=\Delta \mp J$ , $
\varepsilon _{\pm }=\frac{\varepsilon _{1}\pm \varepsilon _{2}}{\sqrt{2}}$.
For the fibre, we define
\begin{equation}
\hat{r}_{\pm k}=\frac{1}{\sqrt{2}}(\hat{r}_{1k}\pm \hat{r}_{2k}),\hat{l}
_{\pm k}=\frac{1}{\sqrt{2}}(\hat{l}_{1k}\pm \hat{l}_{2k}),\hat{d}_{\pm k}=
\frac{\hat{r}_{\pm k}+\hat{l}_{\pm k}}{\sqrt{2}},\hat{c}_{\pm k}=\frac{\hat{r
}_{\pm k}-\hat{l}_{\pm k}}{\sqrt{2}}. \label{bsoutput}
\end{equation}
Thus, $\hat{H}_{f}$ can be rewritten as
\begin{equation}
\hat{H}_{f}=\int_{0}^{\infty }\Delta _{k}dk[\hat{d}_{+k}^{\dagger }\hat{d}
_{+k}+\hat{d}_{-k}^{\dagger }\hat{d}_{-k}+\hat{c}_{+k}^{\dagger }\hat{c}
_{+k}+\hat{c}_{-k}^{\dagger }\hat{c}_{-k}]+\sqrt{2}\xi \int_{0}^{\infty }dk[
\hat{a}_{+}^{\dagger }\hat{d}_{+k}+\hat{a}_{-}^{\dagger }\hat{d}_{-k}+h.c.].
\end{equation}
We see that the cavity modes are decoupled with the fiber mode $\hat{c}_{+k}$
and $\hat{c}_{-k}$. Choosing parameters $\omega _{m}=2J$, we can rewrite {\
Eqs.(\ref{eq-H_int3})} in the frame rotating at $U=exp\{-i\,t[-2J(\hat{a}
_{+}^{\dagger }\hat{a}_{+}+\int_{0}^{\infty }\Delta _{k}dk\hat{d}
_{+k}^{\dagger }\hat{d}_{+k})+\omega _{m}(\hat{b}_{+}^{\dagger }\hat{b}_{+}+
\hat{b}_{-}^{\dagger }\hat{b}_{-})]\}$ and neglect the terms rapidly
oscillating terms, the Hamiltonian\bigskip

\begin{equation}
\hat{H}_{s}=\ \Delta _{-}(\hat{a}_{+}^{\dagger }\hat{a}_{+}+\hat{a}
_{-}^{\dagger }\hat{a}_{-})+\varepsilon _{-}(\hat{a}_{-}^{\dagger }+\hat{a}
_{-})+\frac{g}{\sqrt{2}}(\hat{a}_{+}^{\dagger }\hat{a}_{-}\hat{b}
_{-}^{\dagger }+\hat{a}_{+}\hat{a}_{-}^{\dagger }\hat{b}_{-}),  \label{Hs}
\end{equation}
and
\begin{equation}
\hat{H}_{f}=\int_{0}^{\infty }dk[\Delta _{kJ}\hat{d}_{+k}^{\dagger }\hat{d}
_{+k}+\Delta _{k}(\hat{d}_{-k}^{\dagger }\hat{d}_{-k}+\hat{c}_{+k}^{\dagger }
\hat{c}_{+k}+\hat{c}_{-k}^{\dagger }\hat{c}_{-k})]+\sqrt{2}\xi
\int_{0}^{\infty }dk[\hat{a}_{+}^{\dagger }\hat{d}_{+k}+\hat{a}_{-}^{\dagger
}\hat{d}_{-k}+h.c.].  \label{Hf}
\end{equation}
where $\Delta _{kJ}=\Delta _{k}+2J$. The Hamiltonian {Eq.(\ref{Hs}) }
indicate the three-body interaction between cavities and the oscillator,
which is exact the same with Ref. \cite{Komar2013a} where the nonlinearity
has been analyzed. In a single cavity optomechanical system, the effective
photon-photon interactions $g^{2}/\omega _{m}$ is suppressed by the
condition that the mechanical frequency is much larger than the coupling $g$
, i.e., $\omega _{m}\gg g$, while the three-body interaction (\ref{Hs}) has
its advantage \cite{Komar2013a} that photons in the two optical modes can be
resonantly exchanged by absorbing or emitting a phonon via three-mode
mixing; therefore, the restraint $\omega _{m}\gg g$ can be overcome. Since
our system can be simplified as \cite{Komar2013a} , one can see that the
nonlinearity should be exist and does not restrict by the condition $\omega
_{m}\gg g$. Most importantly, the Hamiltonian {Eq.(\ref{Hs}) and Eq.(\ref{Hf}
) exhibit clearly the conversion between the quasi-mode between }$\hat{a}
_{+}^{{}}$ and $\hat{a}_{-}$ under the witness of $\hat{b}_{-}$ so that we
can realize the exchange between $\hat{d}_{+k}$ and $\hat{d}_{-k}$.
Therefore, with the interaction, we can potentially realize four ports
router.

\subsection*{Photon Blockade}

Now we first investigate the nonlinearity of the photons within the cavity
and temporally forget the coupling with the fibre. The dynamics of the
system obeys the master equation
\begin{equation}
\frac{d\hat{\rho}}{dt}=-i[\hat{H}_{cav}^{^{\prime }}+\hat{H}_{om}^{^{\prime
}},\hat{\rho}]+\sum_{i=1,2}(\hat{L}_{\hat{a}_{i}}+\hat{D}_{\hat{b}
_{i}})\hat{\rho}  \label{master}
\end{equation}
where $\hat{L}_{\hat{a}_{i}}=\kappa (2\hat{a}_{i}\cdot \hat{a}_{i}^{\dag }-
\hat{a}_{i}^{\dag }\hat{a}_{i}\cdot -\cdot \hat{a}_{i}^{\dag }\hat{a}_{i})$,
$\hat{D}_{\hat{b}_{i}}=\gamma _{m}(n_{thm}+1)(2\hat{b}_{i}\cdot \hat{b}
_{i}^{\dag }-\hat{b}_{i}^{\dag }\hat{b}_{i}\cdot -\cdot \hat{b}_{i}^{\dag }
\hat{b}_{i})+\gamma _{m}(2\hat{b}_{i}^{\dag }\cdot \hat{b}_{i}-\hat{b}_{i}
\hat{b}_{i}^{\dag }\cdot -\cdot \hat{b}_{i}\hat{b}_{i}^{\dag })$, $i=1,2$.
To characterize the nonlinearity of optical modes, we employ the equal-time
second-order correlation functions
\begin{equation}
g_{ij}^{(2)}(0)=\frac{\langle {\hat{a}_{i}^{\dagger }\hat{a}_{j}^{\dagger }
\hat{a}_{j}\hat{a}_{i}}\rangle }{\langle {\hat{a}_{i}^{\dagger }\hat{a}_{i}}
\rangle \langle {\hat{a}_{j}^{\dagger }\hat{a}_{j}}\rangle }.  \label{g20e}
\end{equation}
For $i=j$, the function $g_{ii}^{(2)}(0)$ [$g_{jj}^{(2)}(0)$] denotes the
self-correlation, and $g_{ij}^{(2)}(0)$ ($i\neq j)$ express the
cross-correlation. If the correlation function $g_{ij}^{(2)}(0)<1$
we say the photon anti-bunching, and the limit $g_{i}^{(2)}(0)=0$
corresponds to the thorough photon blockade effect, which means that only
one photon can exist, and the another photon will be blockaded.

\begin{figure}[htb]
\centering \includegraphics[width=16cm]{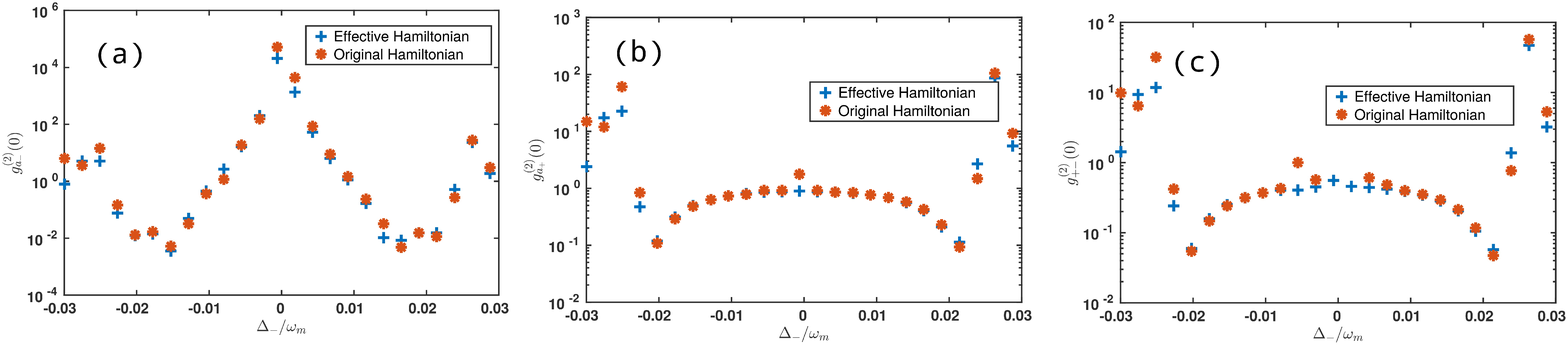}
\caption{(a)Plot a relation between correlation function $g^{(2)}(0)$ of $ a_- $ and detuning
$\Delta_-$, blue dot for solve master equation with effective Hamiltonian  red dot for original
Hamiltonian. (b) Correction $g^{(2)}(0)$ of $a_+$ as function of $\Delta_-$. (c) Cross correlation
function $g^{(2)}_{-+}(0)$ versus  detuning $\Delta_-$.
Other parameters are $\protect\omega_{m_1}=\protect
\omega_{m_2}=\protect\omega_m,$ $J=2\protect\omega_m,g=0.03\protect\omega_m,$
$\protect\kappa=10^{-3}\protect\omega_m,$ $\protect\gamma_m=\protect\kappa
/200,\protect\epsilon_1=1.1 \times 10^{-4}\protect\omega_m,\protect\epsilon
_2=-\protect\epsilon_1$ }
\label{fig2}
\end{figure}

Now, we show the nonlinearity by comparing the numerically solution of the
master equation {Eq.(\ref{master})} with that $\hat{H}_{cav}^{^{\prime }}+
\hat{H}_{om}^{^{\prime }}$ are substituted with effective Hamiltonian Eq.(
\ref{Hs}) where the subscripts $i=1,2$ for the superoperators $\hat{L}_{\hat{
a}_{i}}$ and $\hat{D}_{\hat{b}_{i}}$ are easily changed to $i=-,+$ because
we assume the two cavity modes as well as mechanical modes with equal decay
rate respectively. As shown in {Fig.\ref{fig2}}, we see that the solution
of master equation with the effective Hamiltonian coincides with that of
master equation with original Hamiltonian in most of region except some
unstable date resulted from the cut-off error, which show that the effective
Hamiltonian method is reliable. We will employ the effective Hamiltonian Eq.(
\ref{Hs}) in the calculation of the photon router procession. More
importantly, we observe that $g_{ij}^{(2)}(0)$ ($i,j=-,+$) achieves their
minimum values around $\Delta _{-}=\pm \frac{g}{\sqrt{2}}$, which means that
the system can suppress the simultaneous two-photon creations in any of the
mode $\hat{a}_{-}$ and $\hat{a}_{+}$, especially the cross mode between $
\hat{a}_{-}$ and $\hat{a}_{+}$. That is to say, in the coupled two cavity
optomechanical system, there is most possibly only one photon existence.
Thus, the property can be potentially used as a single photon router if we
can control it. The photon-blockade is resulted from three-body interactions
that lead to destructive interference of optical modes. The conclusion is
also obtained in \cite{Komar2013a} where the destructive interference is
analyzed with eigenstate of the Hamiltonian {Eq.(\ref{Hs})}. \ The
three-body interaction is still dependent on the coupling $g$ see {Eq.(\ref
{Hs}), therefore} the strong coupling strength is still welcome. But the
nonlinearity is not proportional to $\frac{g^{2}}{\omega _{m}}$, which means
that the nonlinearity is not limited by the condition $g\ll \omega _{m}$.

\subsection*{single-photon router}

Quantum router is a hinge device for large-scale network communications. How
to design quantum router arouse a lot of interests\cite
{Aoki2009,Hoi2011,Zhou2013,Lu2014a,Jiang2012b,Lemr2013,Agarwal2012}. To
satisfy the requirements of quantum information, a suitable quantum router
should be worked at single-photon state. Photon blockade effect is a
effective method to realize the single-photon router.

As we have shown in the {Fig.\ref{fig2}}, there is a good photon blockade
phenomenon in this optomechanical system. We can reasonably assume that the
device is only allow a single photon transport. Therefore we will only
consider a single excitation in the system. Now, we employ the two coupled
optomechanical cavities to couple to two waveguide (CRW) shown in {Fig.\ref
{figure_configure}b}. In order to employ the quasi-mode, we introduce medium
as phase shifter and beam splitters to generate the quasi-mode. One can easy
deduce that the four outputs will satisfy the relation {Eq.(\ref{bsoutput})}
. We now calculate the photon number of the four ports. Under the
Hamiltonian {Eq.(\ref{Hs}) and (\ref{Hf})}, the bases is denoted as $
|n_{-},n_{+},n_{b},n_{\hat{d}_{-k}},n_{\hat{d}_{+k}}\rangle $, thus we can
write the wave function with only a single excitation as
\begin{equation}
|\Psi (t)\rangle =\alpha _{-}|1,0,0,0,0\rangle +\alpha _{+}|0,1,1,0,0\rangle
+\int_{0}^{\infty }dk[\mu _{k}|0,0,0,1_{k},0\rangle +\eta
_{k}|0,0,1,0,1_{k}\rangle ], \label{Psi}
\end{equation}
In terms of the left- and right-propagation modes, if we assume a photon
packet is incident onto the cavity from the port $r_{-k}$.
The wave function obey Schrodinger equation with Hamiltonian $\hat{H}=\hat{H}
_{s}+\hat{H}_{f}$. In the long-time limit,we can find the solution of wave
function
\begin{equation*}
|\Psi (t\rightarrow \infty )\rangle =\int_{0}^{\infty }dk[\mu
_{k}(0)e^{-i\Delta _{k}t}(r_{-k}\hat{r}_{-k}+l_{-k}\hat{l}_{-k})+\mu
_{k}^{\prime }(0)e^{-i\Delta _{kJ}t}(r_{+k}\hat{r}_{+k}+l_{+k}\hat{l}
_{+k})]|\emptyset \rangle
\end{equation*}
The details of calculation can be found in part methods. Therefore, the four
output photon number are obtained as
\begin{eqnarray}
N_{r_{-}}^{(out)} &=&\frac{\pi |G_{1}|^{2}}{\epsilon }-2\pi
|G_{1}|^{2}\gamma ^{2}[\frac{1}{\epsilon }\frac{\gamma ^{2}+g^{2}+(\delta
^{\prime }+\epsilon )^{2}}{\mathcal{F}_{++}\mathcal{F}_{+-}\mathcal{F}_{-+}
\mathcal{F}_{--}} \\
&&+\frac{\sqrt{2}}{4g\gamma }(\frac{1}{g/\sqrt{2}+i\gamma }\frac{3g^{2}/2-i
\sqrt{2}g\gamma }{\mathcal{F}_{++}^{\ast }\mathcal{F}_{+-}}+\frac{1}{g/\sqrt{
2}-i\gamma }\frac{3g^{2}/2+i\sqrt{2}}{\mathcal{F}_{-+}^{\ast }\mathcal{F}
_{--}})]  \notag  \label{N_r-}
\end{eqnarray}
\begin{eqnarray}
N_{l_{-}}^{(out)} &=&2\pi |G_{1}|^{2}\gamma ^{2}[\frac{1}{\epsilon }\frac{
\gamma ^{2}+(\delta ^{\prime }+\epsilon )^{2}}{\mathcal{F}_{++}\mathcal{F}
_{+-}\mathcal{F}_{-+}\mathcal{F}_{--}} \\
&&+\frac{\sqrt{2}}{4g\gamma }(\frac{1}{g/\sqrt{2}+i\gamma }\frac{g^{2}/2-i
\sqrt{2}g\gamma }{\mathcal{F}_{++}^{\ast }\mathcal{F}_{+-}}+\frac{1}{g/\sqrt{
2}-i\gamma }\frac{g^{2}/2+i\sqrt{2}}{\mathcal{F}_{-+}^{\ast }\mathcal{F}_{--}
})]  \notag  \label{N_l-}
\end{eqnarray}
\begin{eqnarray}
N_{r_{+}}^{(out)} &=&\pi |G_{1}|^{2}g^{2}\gamma ^{2}[\frac{1}{\epsilon }
\frac{1}{\mathcal{F}_{++}\mathcal{F}_{+-}\mathcal{F}_{-+}\mathcal{F}_{--}} \\
&&+\frac{\sqrt{2}}{4g\gamma }(\frac{1}{g/\sqrt{2}+i\gamma }\frac{1}{\mathcal{
\ F}_{++}^{\ast }\mathcal{F}_{+-}}+\frac{1}{g/\sqrt{2}-i\gamma }\frac{1}{
\mathcal{F}_{-+}^{\ast }\mathcal{F}_{--}})]  \notag \\
N_{l_{+}}^{(out)} &=&N_{r_{+}}^{(out)}  \label{N_r+}
\end{eqnarray}
with $\mathcal{F}_{\pm \pm }=\delta ^{\prime }\pm g/\sqrt{2}\pm \gamma
+i\epsilon $, where $\delta ^{\prime }=\delta -\Delta _{-}$. We can clearly
see that if $g=0$, $N_{l_{+}}^{(out)}=N_{r_{+}}^{(out)}=0$, and $
N_{r_{-}}^{(out)}(N_{l_{-}}^{(out)})\neq 0$, which means that without the
mechanical oscillator we only have two-port router, and the optomechanical
coupling is necessary for us to realize multi-port router.

\begin{figure}[htb]
\centering\includegraphics[width=16cm]{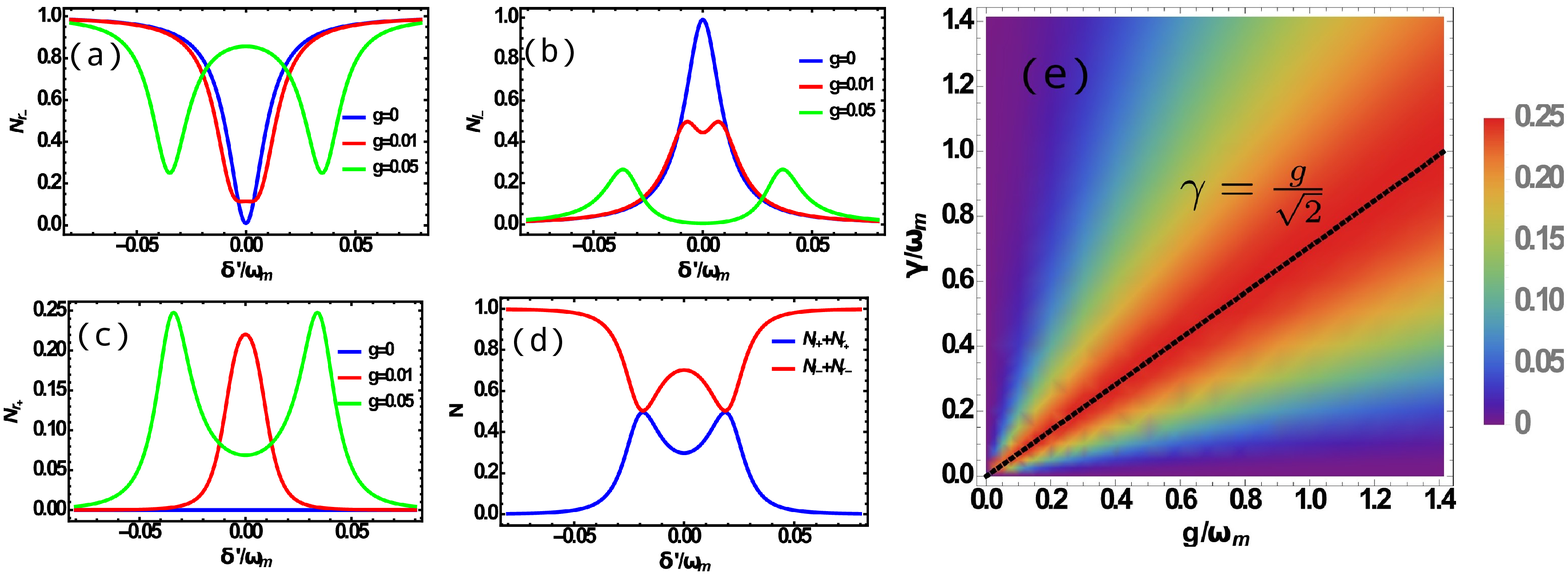}
\caption{ (a),(b),(c) Photon number $N_{r-},N_{l-},N_{r_{+}}$ as function of
$\protect\delta ^{\prime }$ for several values of $g$ where $\protect\gamma
=0.01$. (d) $N_{l_+}+N{r_+}$ and $N_{l_-}+N_{r_-}$ as function of $\protect
\delta^\prime$ with $g=0.04$. It is naturally satisfied normalized condition
$N_{l_+}+N_{r_+}+N_{l_-}+N_{r_-} = 1$. (e)Photon number $N_{l_+}$($N_{r_+})$
versus $\protect\gamma$ and $g$ when $\protect\delta^\prime=0$. And $\protect
\epsilon =0.0001$ all the parameters were normalized by $\protect\omega _{m}$
.}
\label{figure3}
\end{figure}

We plot the output photon number of the four ports as a function of $\delta
^{\prime }$ for several values of $g$ in {Fig.\ref{figure3}(a,b,c)}. If $g=0
$ (means without the coupling of radiation pressure), when $\delta ^{\prime
}=0$ ($\delta =\Delta _{-}$ denotes that the input photon is on resonant
with the cavity fields), the single photon will almost transmit into the
left port $\hat{l}_{-k}$ which was equivalent to a common cavity waveguide coupled
system which present a perfect reflection at resonance region. With the
increasing of $\delta ^{\prime }$, the photon will be partially transmitted
and partially be reflected, but they are still of one peak (valley).
Attributing to three-body interactions $N_{r+},N_{l_{+}}$ occurred when $g>0$
. However, increasing the values of $g$, for example $g=0.05\omega _{m}$,
the single peak (valley) is split into two peaks (valleys) because the
movable mirror participates the three-body interaction so that we can see
the symmetry peaks (valleys). Most importantly, the one port input signal
can be distributed into four ports see Fig.\ref{figure3}(a),(b) and (c),
while for $g=0$, we can receive only two ports signals $N_{l_{-}}$ and $
N_{r_{-}}$. This result has been explained above.  Therefore, with the
assistant of the two coupled cavity optomechanical system, we can realize
multi-port router. We parcel the four-port output into two parts $
N_{l_{+}}+N_{r_{+}},N_{l_{-}}+N_{r_{-}}$ because they denote the difference
whether the optomechanical coupling is included or not. Though we can
transport the photon via the optomechanical coupling, the probability of
transportion $N_{l_{+}}+N_{r_{+}}$ is still less than $N_{l_{-}}+N_{r_{-}}$
under the group of the paramesters. In order to optimized $N_{r_{+}}$($
N_{l_{+}}$), we plot optimized $N_{r_{+}}$($N_{l_{+}}$) as function of the
parameters $g$ and $\gamma $ shown in {Fig.\ref{figure3}(e)}. We observe
that  when there is an optimized value $N_{r_{+}}$($N_{l_{+}}$) along the
line $\gamma =\frac{g}{\sqrt{2}}$, which exhibit that the balance between
cavity waveguide coupling and optomechanical interaction is helpful to the
multi-port router procession.

\section*{Conclusion}

We put forward a scheme to realize multi-port router using two coupled
cavity optomechanical system. We first demonstrate that our system with the
Hamiltonian {Eqs.(\ref{Hs}) can be effectively equal to} the three-body
interaction between cavities and the oscillator which has been shown in \cite
{Komar2013a}. The nonlinearity in the three-mode mixing is not proportion to
$g^{2}/\omega _{m}$ and can overcome the restraint $\omega _{m}\gg g$. We
also numerically show the nonlinearity and correction of the effective
interaction. By coupling the two coupled cavity optomechanical system to
waveguide, we calculate the output photon number of the multi-port router.
Our results show that the presented system can work as multi-port router
under the witness of the optomechanical coupling. Since the two coupled
optomechanical cavity is similar with the experiment \cite{Chang2014} where
the optomechanical coupling is ignored. If the the optomechanical coupling
is strong enough, our scheme should be realizable.

\bigskip

\section*{METHODS}

\subsection*{router}

Now we solve the Schrodinger equation of this system with Hamiltonian $
\hat{H}=\hat{H}_s + \hat{H}_f$ and wave function {Eq.(\ref{Psi})}.
\begin{eqnarray}  \label{diff_equantion}
\dot{\alpha}_{-} &=&-i[\Delta _{-}\alpha _{-}+\frac{g}{\sqrt{2}}\alpha _{+}+
\sqrt{2}\xi \int_{0}^{\infty }dk\mu _{k}],  \notag \\
\dot{\alpha}_{+} &=&-i[\Delta _{-}\alpha _{+}+\frac{g}{\sqrt{2}}\alpha _{-}+
\sqrt{2}\xi \int_{0}^{\infty }dk\eta _{k}],  \notag \\
\dot{\mu}_{k} &=&-i[\Delta _{k}\mu _{k}+\sqrt{2}\xi \alpha _{-}], \\
\dot{\eta}_{k} &=&-i[\Delta _{kJ}\eta _{k}+\sqrt{2}\xi \alpha _{+}].  \notag
\end{eqnarray}
We assume that initially the cavity is in the vacuum state, and a single
photon with the waveguide, i.e., $|0,0,0,1_{k},0\rangle $ is prepared in a
wave packet with a Lorentzian spectrum, the initial condition reads $\mu
_{k}(0)=\frac{G_{1}}{\Delta _{k}-\delta +i\epsilon }$. Using Laplace
transformation, the differential equations {Eq. (\ref{diff_equantion})}
become
\begin{eqnarray}
s\widetilde{\alpha }_{-} &=&-i[\Delta _{-}\widetilde{\alpha }_{-}+\frac{g}{
\sqrt{2}}\widetilde{\alpha }_{+}+\sqrt{2}\xi \int_{0}^{\infty }dk\widetilde{
\mu }_{k}],  \notag \\
s\widetilde{\alpha }_{+} &=&-i[\Delta _{-}\widetilde{\alpha }_{+}+\frac{g}{
\sqrt{2}}\widetilde{\alpha }_{-}+\sqrt{2}\xi \int_{0}^{\infty }dk\widetilde{
\eta }_{k}],  \notag \\
s\widetilde{\mu }_{k}&=&-i[\Delta _{k}\widetilde{\mu }_{k}+ \sqrt{2}\xi
\widetilde{\alpha }_{-}]+\mu _{k}(0),  \notag \\
s\widetilde{\eta }_{k} &=&-i[\Delta _{kJ}\widetilde{\eta }_{k}+\sqrt{2}\xi
\widetilde{\alpha }_{+}],
\end{eqnarray}
where $\gamma _{j}=2\pi \xi _{j}^{2}$ denoting the cavities loss into the
waveguide. If there is no the other decay except the exchange between the
cavities and the waveguide, $\gamma _{j}$ will be equal to the decay rate of
the cavity which we have mentioned in {Fig.\ref{fig2}}.
For simplicity, we consider $\xi _{1}=\xi _{2}$ so that $\gamma =\gamma
_{1}=\gamma _{2}$. In the long-time limit, the coefficients $\mu _{k}(\infty
)$ and $\eta _{k}(\infty )$ are obtained after inverse Laplace
transformation.
\begin{eqnarray}
\mu _{k}(\infty ) &=&\frac{2(\gamma ^{2}+\widetilde{\Delta }_{k}^{2})-g^{2}}{
2(\widetilde{\Delta }_{k}+i\gamma )^{2}-g^{2}}\mu _{k}(0)e^{-i\Delta _{k}t},
\\
\eta _{k}(\infty ) &=&-\frac{2\sqrt{2}ig\gamma }{2(\widetilde{\Delta }
_{kJ}+i\gamma )^{2}-g^{2}}\mu _{k}^{\prime }(0)e^{-i\Delta _{kJ}t}.  \notag
\end{eqnarray}
In terms of the left- and right-propagation modes, if we assume a photon
packet is incident onto the cavity from the port $r_{-k}$, then the initial
state can be written as

\begin{equation}
|\Psi (0)\rangle =\int_{0}^{\infty }dk\mu _{k}(0)\hat{r}_{-k}|\emptyset
\rangle =\frac{1}{\sqrt{2}}\int_{0}^{\infty }dk\mu _{k}(0)(\hat{d}_{-k}+\hat{
c}_{-k})|\emptyset \rangle ,  \label{M0}
\end{equation}
which means that the single photon input from the port $r_{-k}$ can be
considered as a superposition between a quasiparticle $\hat{d}_{-k}$ and a
quasiparticle $\hat{c}_{-k}$. In the long-time limit, the wave function
becomes under the Hamiltonian Eq.(\ref{Hs}) and Eq.(\ref{Hf})

\begin{equation*}
|\Psi (t\rightarrow \infty )\rangle =\int_{0}^{\infty }dk[\mu
_{k}(0)e^{-i\Delta _{k}t}(r_{-k}\hat{r}_{-k}+l_{-k}\hat{l}_{-k})+\mu
_{k}^{\prime }(0)e^{-i\Delta _{kJ}t}(r_{+k}\hat{r}_{+k}+l_{+k}\hat{l}
_{+k})]|\emptyset \rangle
\end{equation*}
,
where the first bracket with the factor $e^{-i\Delta _{k}t}$ can survive
without Hamiltonian Eq.(\ref{Hs}), while the second bracket with the factor $
e^{-i\Delta _{kJ}t}$ survive only under the condition Eq.(\ref{Hs})
existence. In other words, the photon on the ports $r_{-k}$ and $l_{-k}$ can
be detected even without the mechanical mode, however, if we would like to
obtain photon on the port $r_{+k}$ and $l_{+k}$, the coupling between the
mechanical mode and cavity field is necessary. Then we obtain
\begin{eqnarray}
r_{-k} &=&\sqrt{2}\frac{(\gamma ^{2}+\widetilde{\Delta }_{k}^{2})+(
\widetilde{\Delta }_{k}+i\gamma )^{2}-g^{2}}{2(\widetilde{\Delta }
_{k}+i\gamma )^{2}-g^{2}}, \\
l_{-k} &=&\sqrt{2}\frac{\gamma ^{2}+\widetilde{\Delta }_{k}^{2}-(\widetilde{
\Delta }_{k}+i\gamma )^{2}}{2(\widetilde{\Delta }_{k}+i\gamma )^{2}-g^{2}},
\notag \\
l_{+k} &=&-\frac{2ig\gamma }{2(\Delta _{k}+i\gamma )^{2}-g^{2}},  \notag \\
r_{+k} &=&l_{+k},  \notag
\end{eqnarray}
and the output photon number in {Eq.(\ref{N_r-}),(\ref{N_l-}),(\ref{N_r+})}.

\section*{Acknowledgements (not compulsory)}

We would like to thank Wei-bin Yan for helpful discussions. This work was supported by the NSF of
China under Grant No. 11474044.

\section*{Author contributions statement}
X.L. and L.Z. designed the research, X.L. did the analytic calculations, W.Z.Z. provided numerical 
and prepared figures, L.Z. revised the manuscript and provided overall theoretical support.

\section*{Additional information}
Competing financial interests: The authors declare no competing financial interests.
\end{document}